\documentclass[twocolumn,prb,showpacs,floatfix]{revtex4}

\usepackage{graphicx}
\usepackage{bm}
\usepackage{color}
\usepackage{amsmath}
\usepackage{amsfonts}
\usepackage{amssymb}
\usepackage{epstopdf}

\begin{document}

\preprint{APS/123-QED}

\title{Quantum phase transition between disordered and ordered states in the spin-$\frac{1}{2}$ kagome lattice antiferromagnet 
(Rb$_{1-x}$Cs$_{x}$)$_2$Cu$_3$SnF$_{12}$}

\author{Kazuya Katayama}
\email{katayama.k.ad@m.titech.ac.jp}
\author{Nobuyuki Kurita}
\email{kurita.n.aa@m.titech.ac.jp}
\author{Hidekazu Tanaka}
\email{tanaka@lee.phys.titech.ac.jp}

\affiliation{
Department of Physics, Tokyo Institute of Technology, Oh-okayama, Meguro-ku, Tokyo 152-8551, Japan\\
}
\date{\today}

\begin{abstract}
We have systematically investigated the variation of the exchange parameters and the ground state in the $S\,{=}\,1/2$ kagome-lattice antiferromagnet (Rb$_{1-x}$Cs$_{x}$)$_2$Cu$_3$SnF$_{12}$, via magnetic measurements using single crystals. One of the parent compounds, Rb$_2$Cu$_3$SnF$_{12}$, which has a distorted kagome lattice accompanied by four sorts of nearest-neighbor exchange interaction, has a disordered ground state described by a pinwheel valence-bond-solid state. The other parent compound, Cs$_2$Cu$_3$SnF$_{12}$, which  has a uniform kagome lattice at room temperature, has an ordered ground state with the $q\,{=}\,0$ spin structure. The analysis of magnetic susceptibilities shows that with increasing cesium concentration $x$, the exchange parameters increase with the tendency to be uniform. It was found that the ground state is disordered for $x\,{<}\,0.53$ and ordered for $x\,{>}\,0.53$. The pseudogap observed for $x\,{<}\,0.53$ and the N\'{e}el temperature for $x\,{>}\,0.53$ approach zero at $x_{\rm c}\,{\simeq}\,0.53$. This is indicative of the occurrence of a quantum phase transition at $x_{\rm c}$.
\end{abstract}

\pacs{75.10.Jm, 75.10.Kt, 75.40.Cx}

\maketitle

\section{Introduction}

Geometrically frustrated quantum antiferromagnets, a research frontier in condensed matter physics, have been attracting growing attention owing to the potential realization of exotic ground states such as the spin liquid state.\cite{Balents_Nature2010} One of the simplest and most intriguing frustrated magnets is a Heisenberg antiferromagnet on the kagome lattice (KLAF) composed of corner-sharing triangles. For the classical Heisenberg KLAF with the nearest-neighbor exchange interaction, the ground state is infinitely degenerate owing to the local flexibility of the configuration of the 120$^{\circ}$ spin structure characteristic of the kagome lattice. In the case of a non-classical Heisenberg spin with a large spin quantum number, it has been predicted that the so-called $\sqrt{3}\,{\times}\,\sqrt{3}$ structure is stabilized by the order-by-disorder mechanism.\cite{Chubukov, Sachdev} The most intriguing case is the spin-$\frac{1}{2}$ case, where a noteworthy synergistic effect of the geometric frustrations, the local flexibility, and the quantum fluctuations is expected. A long theoretical debate has reached the consensus that the quantum-disordered state is more stable than any ordered state. However, the nature of the ground state has not been theoretically elucidated. Recent theory suggests the nonmagnetic ground states, such as the valence-bond-solid\,\cite{Lawler_PRL2008,Singh,Sorensen_PRB2009,Evenbly,Hwang1} and quantum spin-liquid states,\cite{Hastings_PRB2000,Wang,Hermele} which are described by a static array of singlet dimers and the superposition of various configurations composed of singlet dimers, respectively.
The nature of low-energy excitations also remains unresolved. The presence of a gap for the triplet excitation is still controversial.\cite{Yan_Science2011,Waldmann_EPJB1998,Depenbrock,Nishimoto_NatComm2013,Nakano_JPSJ2011}

On the  experimental side, considerable effort has been made to search for materials that closely approximate the spin-$\frac{1}{2}$ Heisenberg KLAF.\cite{Hiroi_JPSJ2001,Yoshida_NatComm2012,Okamoto_JPSJ2009,Yoshida_JPSJ2013,Shores_JACS2005,Mendels_PRL2007,Helton,Mueller} However, the materials investigated, many of which are natural minerals, have individual problems such as spatial anisotropy of the exchange network\cite{Yoshida_NatComm2012,Janson}, exchange disorder due to ion substitution\cite{Bert_PRB2007} and lattice distortion due to a structural phase transition.\cite{Reisinger} For these reasons, there has been little clear experimental evidence demonstrating the nature of the ground state and the excitations for the spin-$\frac{1}{2}$ Heisenberg KLAF.


\begin{figure*}[thb]
\begin{center}
\includegraphics[width=0.80\linewidth]{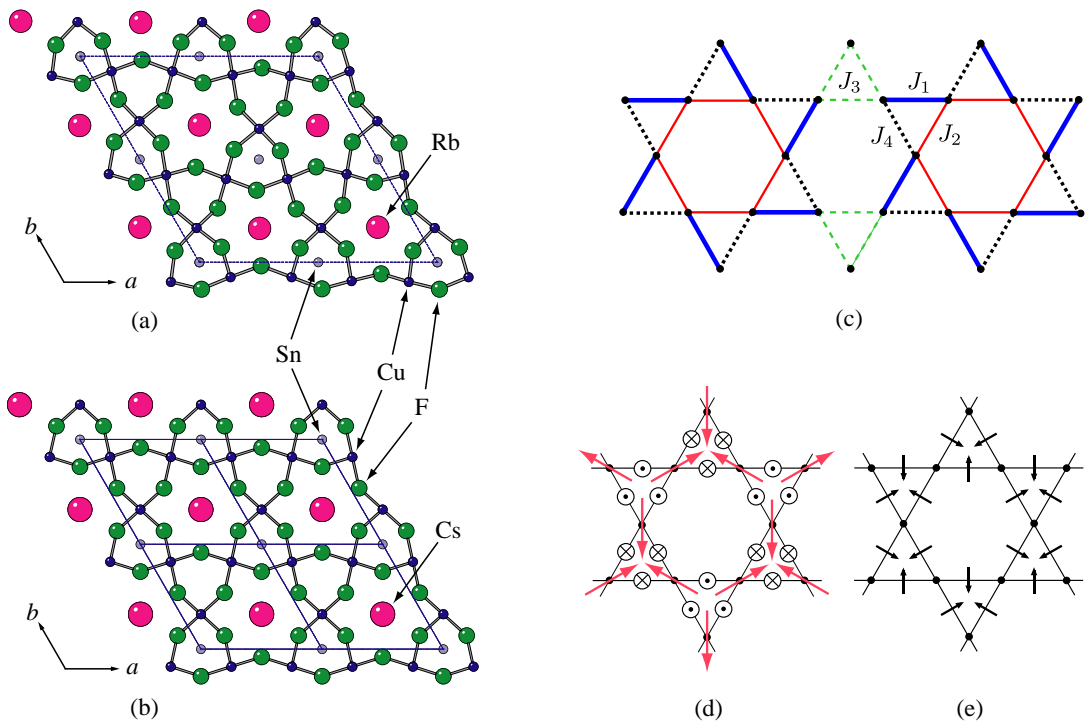}
\end{center}
\caption{Crystal structures of (a) Rb$_2$Cu$_3$SnF$_{12}$ and (b) Cs$_2$Cu$_3$SnF$_{12}$ viewed along the $c$ axis, where F$^-$ ions located outside the kagome layer are omitted. Thin dotted lines denote the chemical unit cells. (c) Configuration of the exchange interactions $J_i$ ($i\,{=}\,1-4$) for Rb$_2$Cu$_3$SnF$_{12}$. Arrangement of the ${\bm D}$ vectors of the DM interaction: (d) $c$-axis component $D^{\parallel}$ and (e) $c$-plane component $D_{\perp}$. The circled dots and circled crosses in (d) and the arrows in (e) represent the local positive directions of the parallel and perpendicular components $D^{\parallel}$ and $D^{\perp}$, respectively. The large arrows in (d) indicate the $q\,{=}\,0$ structure stabilized by the DM interaction.
} 
\label{structure}
\end{figure*}

The cupric fluoride kagome family, $A_2$Cu$_3$SnF$_{12}$ ($A$\,=\,Rb and Cs), which has a trigonal structure, is promising for the comprehensive study of spin-$\frac{1}{2}$ KLAFs.\cite{Morita_JPSJ2008,Ono_PRB2009} Figures\,\ref{structure}(a) and (b) show the crystal structures of Rb$_2$Cu$_3$SnF$_{12}$ and Cs$_2$Cu$_3$SnF$_{12}$ viewed along the $c$ axis, respectively. CuF$_6$ octahedra are linked by sharing their corners in the crystallographic $ab$ plane. Magnetic Cu$^{2+}$ ions with spin-$\frac{1}{2}$ form a kagome lattice in the $ab$ plane. The octahedra are elongated along the principal axes, which is approximately parallel to the $c$ axis owing to the Jahn-Teller effect. Hence, the hole orbitals $d(x^2\,{-}\,y^2)$ of Cu$^{2+}$ are spread in the kagome layer. This leads to a strong superexchange interaction in the kagome layer and a negligible superexchange interaction between layers. 

At room temperature, Rb$_2$Cu$_3$SnF$_{12}$ has a $2a\,{\times}\,2a$ enlarged chemical unit cell in the $ab$ plane, as shown in Fig.\,\ref{structure}(a); thus, the kagome lattice in Rb$_2$Cu$_3$SnF$_{12}$ is not uniform.\cite{Morita_JPSJ2008} There are four sorts of nearest-neighbor exchange interactions as depicted in Fig.\,\ref{structure} (c).\cite{Morita_JPSJ2008} Cs$_2$Cu$_3$SnF$_{12}$ has a uniform kagome lattice at room temperature.\cite{Ono_PRB2009} As the temperature decreases, Cs$_2$Cu$_3$SnF$_{12}$ undergoes a structural phase transition from the trigonal structure to the monoclinic structure at $T_{\rm t}$\,=\,184 K,\cite{Downie} which is closely related to the trigonal structure with a $2a\,{\times}\,2a$ enlarged unit cell.\cite{Ono_PRB2009,Ono_JPSJ2014} 

Since high-purity and sizable single crystals are obtainable, the magnetic properties of these two compounds have been probed in detail by magnetic, neutron scattering and NMR measurements.\cite{Ono_JPSJ2014,Matan_Nature_Phys2010,Matan_PRB2014,Grbic_PRL2013} The magnetic ground state of Rb$_2$Cu$_3$SnF$_{12}$ is a spin singlet with an excitation gap ${\Delta}/k_{\rm B}$ of 27 K.\cite{Morita_JPSJ2008,Matan_Nature_Phys2010,Matan_PRB2014,Grbic_PRL2013} Neutron inelastic scattering experiments revealed that the ground state is the pinwheel valence-bond-solid (VBS) state, in which singlet dimers are situated on the strongest exchange interaction $J_1$ shown in Fig.\,\ref{structure}(c).\cite{Matan_Nature_Phys2010,Matan_PRB2014,Yang_PRB2009,Khatami_PRB2011} The gapped ground state in Rb$_2$Cu$_3$SnF$_{12}$ arises from the inequivalence of the exchange interactions, i.e., $J_1/k_{\rm B}\,{=}\,216$ K, $J_2\,{=}\,0.95J_1$, $J_3\,{=}\,0.85J_1$ and $J_4\,{=}\,0.55J_1$.\cite{Matan_Nature_Phys2010} 

On the other hand, Cs$_2$Cu$_3$SnF$_{12}$ exhibits a magnetic ordering at $T_{\rm N}\,{=}\,20.0$ K.\cite{Ono_PRB2009} In the ordered phase, the so-called $q\,{=}\,0$ spin structure is realized.\cite{Ono_JPSJ2014} The effect of the antisymmetric Dzyaloshinsky-Moriya (DM) type interaction on the ground state was investigated numerically by C\'{e}pas \textit{et al}.,\cite{Cepas} who assumed the same configuration of the ${\bm D}$ vector as that in Cs$_2$Cu$_3$SnF$_{12}$ [Fig.\,\ref{structure}(d)]. They demonstrated that with increasing longitudinal component $D^{\parallel}$, the disordered state changes at $(D^{\parallel}/J)_{\rm c}\,{\approx}\,0.1$ to the ordered state with the $q\,{=}\,0$ structure. The magnitude of the ${\bm D}$ vector in A$_2$Cu$_3$SnF$_{12}$ was evaluated to be $D^{\parallel}/J\,{\simeq}\,1/4$ from the analyses of the dispersion relations.\cite{Matan_Nature_Phys2010,Ono_JPSJ2014} Thus, the magnetic ordering observed in Cs$_2$Cu$_3$SnF$_{12}$ can be attributed to the large DM interaction. Although the ground state of Cs$_2$Cu$_3$SnF$_{12}$ is ordered, a noteworthy quantum many-body effect on the spin wave excitations was observed.\cite{Ono_JPSJ2014} The excitation energies are markedly renormalized downward with respect to the linear spin-wave result in contrast to the conventional quantum renormalization, in which the excitation energies are renormalized upward.\cite{dCP,Endoh,Igarashi,Singh2,Ronnow}  

In Rb$_2$Cu$_3$SnF$_{12}$, the lowest excitation is located at the ${\Gamma}$ point, although the lowest excitation is expected to be located at the $K$ point within the Heisenberg type exchange interaction. This is because the large DM interaction splits the triply degenerate triplet excitations into two levels, $S^z\,{=}\,0$ and $S^z\,{=}\,{\pm}1$ branches, and the energy of the $S^z\,{=}\,{\pm}1$ branch is minimized at the ${\Gamma}$ point with increasing magnitude of $D^{\parallel}$.\cite{Matan_Nature_Phys2010,Hwang2} If the inequivalence of the exchange interactions becomes small, it is expected that the gap at the ${\Gamma}$ point will decrease and a transition from the singlet ground state to the ordered ground state will occur. The exchange interactions in Cs$_2$Cu$_3$SnF$_{12}$ are similar to those in the uniform case.\cite{Ono_JPSJ2014} Thus, we can expect a quantum phase transition in (Rb$_{1-x}$Cs$_{x}$)$_2$Cu$_3$SnF$_{12}$ upon varying the cesium concentration $x$. This motivated us to investigate the magnetic properties of (Rb$_{1-x}$Cs$_{x}$)$_2$Cu$_3$SnF$_{12}$.

In this paper, we present the results of magnetic measurements of (Rb$_{1-x}$Cs$_{x}$)$_2$Cu$_3$SnF$_{12}$ with various $x$ values and the specific-heat measurement of Cs$_2$Cu$_3$SnF$_{12}$. The analysis of magnetic susceptibility using exact diagonalization calculations for a 12-site kagome cluster shows that the exchange interactions tend to become more uniform as $x$ increases. It was found that with increasing $x$, the disordered ground state changes to the ordered state at $x_{\rm c}\,{\simeq}\,0.53$, as shown below. The magnitude of the spin gap decreases and approaches zero at $x_{\rm c}$, while the ordering temperature $T_{\rm N}$ decreases as $x$ decreases from 1 and also appears to be zero at $x_{\rm c}$. These observations indicate the occurrence of a quantum phase transition at $x_{\rm c}\,{\simeq}\,0.53$.

\begin{figure}[t]
 \begin{center}
\includegraphics[width=0.95 \linewidth]{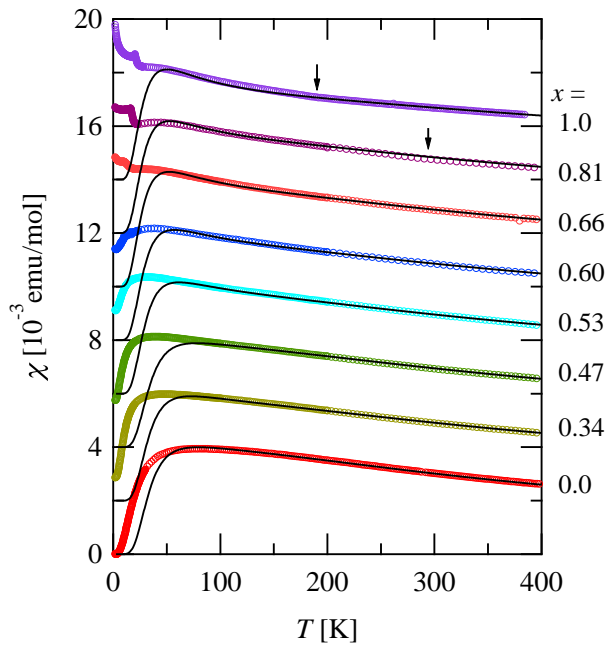}
\end{center}
\caption{
Temperature dependence of magnetic susceptibility of (Rb$_{1-x}$Cs$_{x}$)$_2$Cu$_3$SnF$_{12}$ measured at $H\,{=}\,1$ T for $H\,{\parallel}\,c$ for various $x$. The susceptibility data are shifted upward by multiples of  $2\,{\times}\,10^{-3}$ emu/mol. The data for $x \le 0.47$ are corrected for the Curie-Weiss term attributable to impurities. Arrows indicate anomalies associated with structural phase transitions. Solid curves are fits using the theoretical susceptibilities obtained from the exact diagonalization of a 12-site kagome cluster (see text).
} \label{sus}
\end{figure}

\section{Experimental details}

We synthesized (Rb$_{1-x}$Cs$_{x}$)$_2$Cu$_3$SnF$_{12}$ single crystals from a melt comprising a mixture of Rb$_2$Cu$_3$SnF$_{12}$ and Cs$_2$Cu$_3$SnF$_{12}$ in the ratio of $1\,{-}\,x$ to $x$. Single crystals of $A_2$Cu$_3$SnF$_{12}$ ($A$\,{=}\,Cs, Rb) were grown by a procedure similar to that reported in previous papers.\cite{Morita_JPSJ2008,Ono_PRB2009,Matan_PRB2014,Ono_JPSJ2014} The cesium concentration $x$ was determined by inductively coupled plasma mass spectroscopy (ICP-MS) at the Center for Advanced Materials Analysis, Tokyo Institute of Technology.
We confirmed that the x-ray powder diffraction pattern obtained using MiniFlexI\hspace{-1pt}I (Rigaku) changes systematically with $x$ and that the Bragg peaks are as sharp as those in pure cases with $x\,{=}\,0$ and 1, which indicates high homogeneity of the single crystals.
Magnetic susceptibilities were measured under a magnetic field of 1 T in the temperature range $1.8-400$ K using a superconducting quantum interference device magnetometer (Quantum Design: MPMS XL). Magnetic fields were applied parallel and perpendicular to the $c$ axis. The magnetic susceptibility and magnetization data presented in this paper are corrected for the diamagnetism of core electrons\,\cite{Selwood_Wiley1956} and the Van Vleck paramagnetism, as described in Ref. \onlinecite{Ono_PRB2009}. The specific heat was measured down to 0.36 K in zero magnetic field using a physical property measurement system (Quantum Design: PPMS) by the relaxation method.

\section{Results and Discussion}

\begin{figure}[t]
\begin{center}
\includegraphics[width=0.98\linewidth]{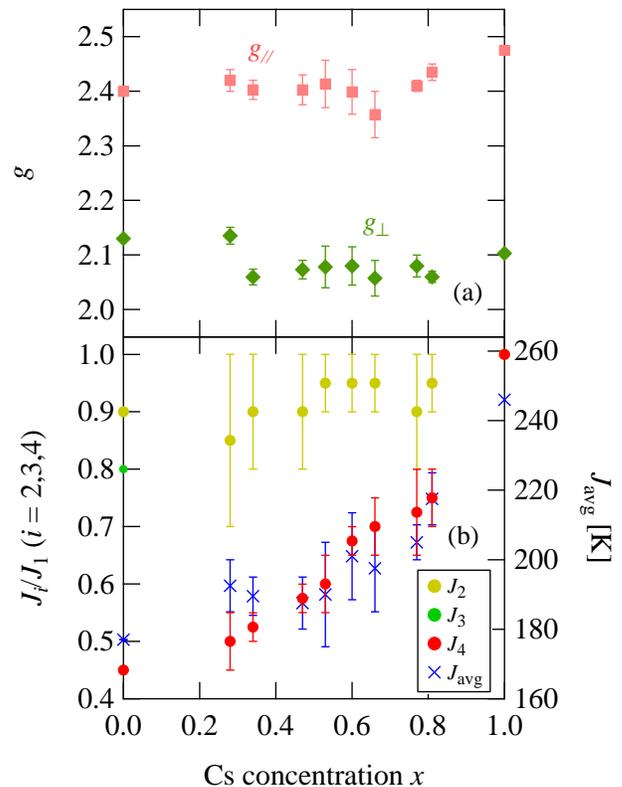}
\end{center}
\caption{
$g$ factors and exchange parameters as a function of the cesium concentration $x$ in (Rb$_{1-x}$Cs$_{x}$)$_2$Cu$_3$SnF$_{12}$, evaluated from the analyses of magnetic susceptibilities. 
(a) $g_{\parallel}$ and $g_{\perp}$ for $H\parallel c$ and $H\perp c$, respectively. 
(b) The individual exchange parameters normalized by $J_1$ and the average exchange interaction $J_{\mathrm avg}$.
} 
\label{fitting_value}
\end{figure}

\begin{figure*}[t]
\begin{center}
\includegraphics[width=0.77\linewidth]{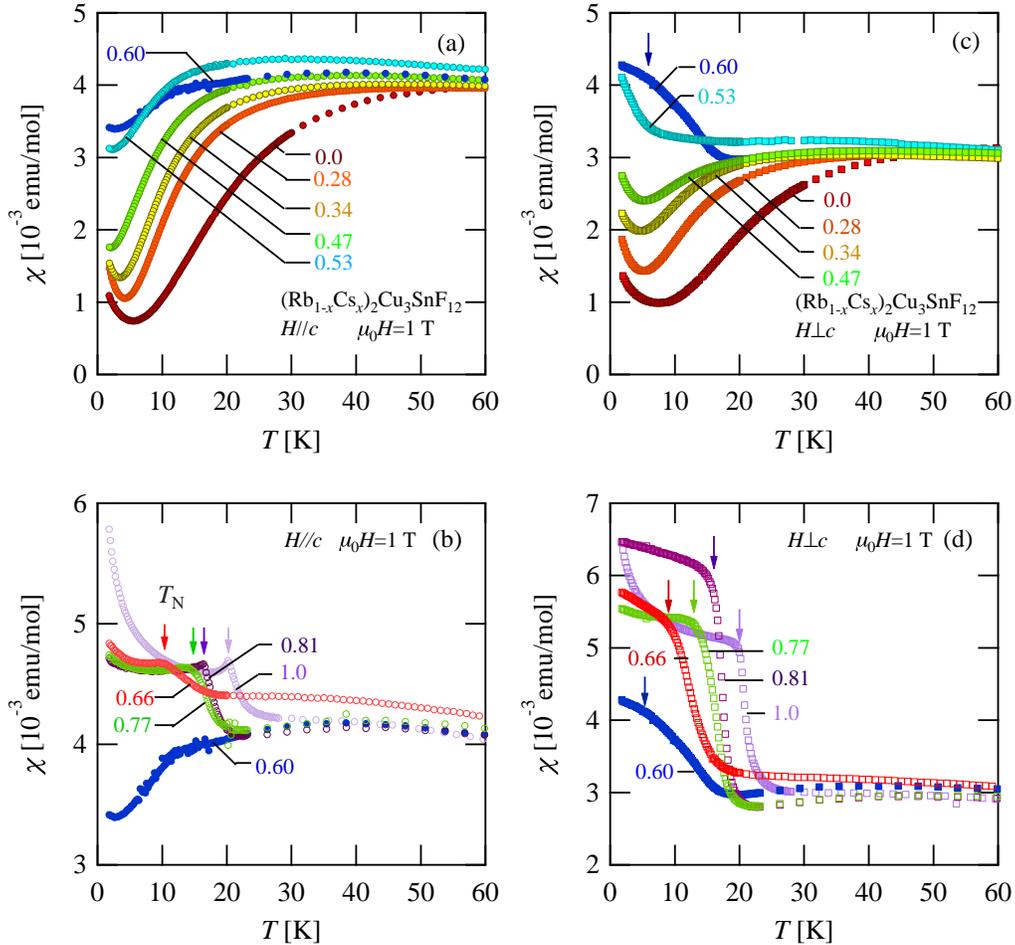}
\end{center}
\caption{
Measured low-temperature magnetic susceptibility of (Rb$_{1-x}$Cs$_{x}$)$_2$Cu$_3$SnF$_{12}$ for various $x$ measured at $H\,=\,1$ T (a) and (b) for $H\,{\parallel}\,c$ and (c) and (d) for $H\,{\perp}\,c$. (a) and (c) show the data for $0\,{\le}\,x\,{\le}\,0.60$, and (b) and (d) show the data for $0.60\,{\le}\,x\,{\le}\,1.0$. Arrows indicate the ordering temperature $T_{\rm N}$.
} 
\label{low_temp}
\end{figure*}

\begin{figure}[h]
\begin{center}
\includegraphics[width=0.9\linewidth]{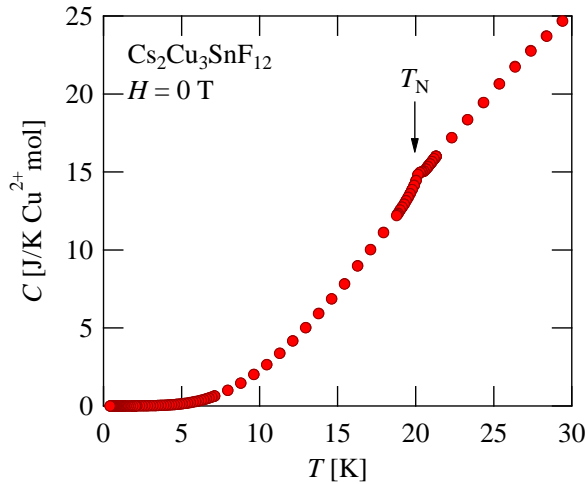}
\end{center}
\caption{Specific heat of Cs$_2$Cu$_3$SnF$_{12}$ as a function of temperature measured at zero magnetic field. The arrow indicates the ordering temperature $T_{\rm N}$.} 
\label{heat}
\end{figure}

\subsection{Exchange parameters}
Figure \ref{sus} shows the temperature dependence of the magnetic susceptibility  $\chi$ of (Rb$_{1-x}$Cs$_{x}$)$_2$Cu$_3$SnF$_{12}$ measured at $H\,{=}\,1$\,T for $H\,{\parallel}\,c$ for various $x$. The data for $x \le 0.47$ are corrected for the Curie-Weiss term attributable to impurities. With decreasing temperature, the susceptibility for Rb$_2$Cu$_3$SnF$_{12}$ ($x\,{=}\,0$) exhibits a rounded maximum at approximately $T_{\rm max}\,{=}\,70$ T and decreases to zero, indicating a gapped singlet ground state. As the cesium concentration $x$ is increased, $T_{\rm max}$ decreases and the magnetic susceptibility has a finite value at $T\,{=}\,0$. With further increasing $x$, a kink anomaly indicating magnetic ordering is observed. Details of the low-temperature susceptibility will be discussed later. For Cs$_2$Cu$_3$SnF$_{12}$ ($x\,{=}\,1.0$), the small bend anomaly shown by an arrow, indicating a structural phase transition, was observed at $T_{\rm t}\,{=}\,184$ K, as previously reported.\cite{Ono_PRB2009} For $x\,{=}\,0.81$, the bend anomaly due to the structural phase transition occurs at $T_{\rm t}\,{=}\,295$ K. This indicates that $T_{\rm t}$ increases with decreasing $x$. 

Assuming that the configuration of the nearest-neighbor exchange interaction shown in Fig.\,\ref{structure}(c) is common to all (Rb$_{1-x}$Cs$_{x}$)$_2$Cu$_3$SnF$_{12}$ on average, we evaluate the exchange parameters $J_i$ $(i=1-4)$ using the exact diagonalization of a 12-site kagome cluster under a periodic boundary condition. For Cs$_2$Cu$_3$SnF$_{12}$ ($x\,{=}\,1$) we assume a uniform kagome lattice for simplification as in a previous paper.\cite{Ono_PRB2009} The calculation procedure has been presented in the previous papers.\cite{Morita_JPSJ2008,Ono_PRB2009} From the fitting to the high-temperature magnetic susceptibility for $T\,{>}\,200$ K, we evaluate the average exchange interaction $J_{\rm avg}$. 
For Rb$_2$Cu$_3$SnF$_{12}$ ($x\,{=}\,0$), we confirmed that the magnetic susceptibility is satisfactorily  reproduced using $J_1/k_{\rm B}\,{=}\,216$ K, $J_2\,{=}\,0.95J_1$, $J_3\,{=}\,0.85J_1$ and $J_4\,{=}\,0.55J_1$ obtained from the analysis of the dispersion relations.\cite{Matan_Nature_Phys2010} 
Thus, individual values of $J_i/J_1$ can be estimated from the fitting to the low-temperature magnetic susceptibility for $T\, <\, 200$ K.
For the intermediate compounds, we found that the calculated susceptibility is only slightly sensitive to $J_3$ for $0.5\,{\leq}\,J_3/J_1\,{\leq}\,1.0$, hence we estimated only $J_2/J_1$ and $J_4/J_1$. Meanwhile, $J_3/J_1$ could not be determined uniquely. 
The solid lines in Fig.\,\ref{sus} are fits with the parameters shown in Fig.\,\ref{fitting_value}. We also performed the same analysis on the susceptibility data for $H\,{\perp}\,c$. The exchange parameters shown in Fig.\,\ref{fitting_value} are obtained by fitting for both $H\,{\parallel}\,c$ and $H\,{\perp}\,c$. In the present analysis, we neglect the DM interaction because its effect on the susceptibility for $T\,{>}\,60$ K is small.\cite{Ono_JPSJ2014} 

As shown in Fig.\,\ref{sus}, the calculated susceptibilities accurately reproduce the experimental susceptibilities for $T\,{>}\,60$ K in all (Rb$_{1-x}$Cs$_{x}$)$_2$Cu$_3$SnF$_{12}$, while for $T\,{<}\,60$ K, the calculated susceptibility decreases more rapidly than the experimental susceptibility. This should be due to the finite-size effect. 
Figure~\ref{fitting_value} summarizes the $x$ dependence of the magnetic parameters for (Rb$_{1-x}$Cs$_{x}$)$_2$Cu$_3$SnF$_{12}$ determined from the susceptibility analyses using the exact diagonalization calculations. 
$g_{\parallel}$ and $g_{\perp}$ denote the $g$ factors for $H\,{\parallel}\,c$ and $H\,{\perp}\,c$, respectively. 
We incorporate the $g$ factor into the fitting parameters because it is difficult to determine the $g$ factor by the usual electron paramagnetic resonance because of the extremely large line-width arising from the large DM interaction. The $g$ factors obtained with the present analysis are independent of $x$. The magnitude of the $g$ factors, i.e., $g_{\parallel}=2.4-2.5$ and $g_{\perp}=2.1$, are consistent with those for K$_2$CuF$_4$ and Rb$_2$CuF$_4$.\cite{Sasaki_JPSJ1995} As shown Fig.\,\ref{fitting_value}(b), the average of the four sorts of exchange interactions $J_{\rm avg}$ increases monotonically as $x \rightarrow 1$. $J_4/J_1$ increases rapidly with increasing $x$. The calculated susceptibility is insensitive to $J_3$ for $0.5\,{\leq}\,J_3/J_1\,{\leq}\,1.0$, as mentioned above. Because the smallest $J_4/J_1$ increases with increasing $x$, we infer that all the exchange interactions approach a uniform value for $x \rightarrow 1$. 

\subsection{Ground-state phase diagram}
Figure \ref{low_temp} shows the low-temperature magnetic susceptibility of (Rb$_{1-x}$Cs$_{x}$)$_2$Cu$_3$SnF$_{12}$ measured at $H\,{=}\,1$ T for $H\,{\parallel}\,c$ and $H\,{\perp}\,c$ for various $x$. For $x\,{\le}\,0.47$, the susceptibility exhibits a small upturn below 7 K, which should be due mainly to the impurity phase. With increasing $x$, the temperature $T_{\rm max}$ giving the rounded maximum of susceptibility decreases. This behavior of susceptibility is considered to be related to the fact that the exchange interactions become uniform with increasing $x$. The low-temperature susceptibility for $x\,{\le}\,0.47$ corrected for the upturn below 7 K shows exponential temperature dependence indicative of the presence of an excitation gap. For $x\,{\le}\,0.53$, no anomaly indicative of magnetic ordering is observed. This shows that the ground state is disordered for $x\,{\le}\,0.53$. On the other hand, the susceptibility for $x\,{\ge}\,0.60$ exhibits kink or bend anomalies, which are suggestive of magnetic ordering. For $x\,{=}\,0.60$, we assigned the temperature at which a small bend anomaly appears for $H\,{\perp}\,c$ as the ordering temperature $T_{\rm N}$. 

Figure \ref{heat} shows the temperature dependence of the specific heat for Cs$_2$Cu$_3$SnF$_{12}$. A tiny cusp anomaly owing to magnetic ordering is observed at $T_{\rm N}\,{=}\,20.0$ K. This ordering temperature coincides with that assigned from the anomaly in the susceptibility. The very small anomaly in the specific heat around $T_{\rm N}$ indicates that little entropy remains for magnetic ordering because of the well-developed short-range spin correlation caused by the large exchange interaction of $J/k_{\rm B}\,{\simeq}\,240$ K and good two-dimensionality. For $x\,{\neq}\,1$, the specific heat anomaly is so small that it is difficult to detect the magnetic ordering.

\begin{figure}[t]
\begin{center}
\includegraphics[width=0.95\linewidth]{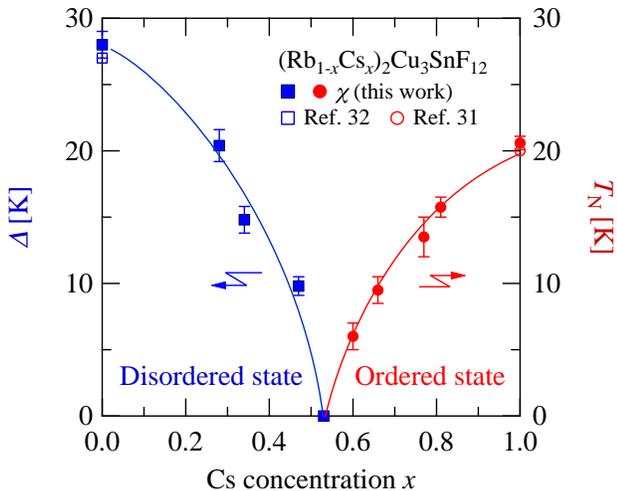}
\end{center}
\caption{
Phase diagram of the spin gap $\Delta$ and N\'eel temperature $T_{\rm N}$ 
for (Rb$_{1-x}$Cs$_{x}$)$_2$Cu$_3$SnF$_{12}$,
determined via magnetic measurements with $H\parallel c$. 
$\Delta$ is estimated by fitting the susceptibility data using Eq.\,(\ref{Gap}). 
Results obtained in other measurements\cite{Matan_Nature_Phys2010, Ono_JPSJ2014} are also shown for comparison.
} 
\label{phase}
\end{figure}

The transition data obtained from the low-temperature susceptibilities are summarized in Fig.\,\ref{phase}. With decreasing $x$, the ordering temperature $T_{\rm N}$ decreases, and $T_{\rm N}$ reaches zero at $x_{\rm c}\,{\simeq}\,0.53$. 

We analyze the low-temperature susceptibility for $x\,{\le}\,0.47$ using the following formula:
\begin{equation}
{\chi}(T) = \frac{C}{T-\Theta} + A \exp{\left(-\frac{\Delta}{k_{\text B}T} \right)} + {\chi}_{0},
\label{Gap}
\end{equation} 
where the first term is the Curie-Weiss term, the second term represents the low-temperature susceptibility for two-dimensional systems with an excitation gap $\Delta$,\cite{Troyer_PRB1994,Stone_PRB2001} and the last constant term arises from the finite susceptibility component in the ground state. 

\begin{figure}[t]
\begin{center}
\includegraphics[width=0.85\linewidth]{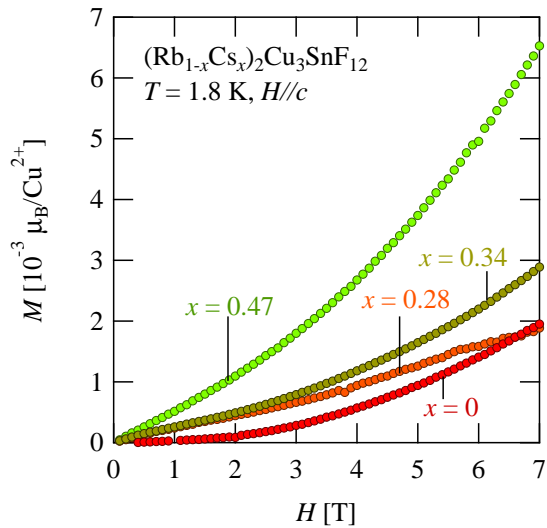}
\end{center}
\caption{
Magnetization curves of (Rb$_{1-x}$Cs$_{x}$)$_2$Cu$_3$SnF$_{12}$ measured for $H\,{\parallel}\,c$ at $T\,=\,1.8\,\rm{K}.$ 
The data were corrected for impurity contributions represented by the Brillouin function. 
} \label{field}
\end{figure}

\begin{figure}[t]
\begin{center}
\includegraphics[width=0.85\linewidth]{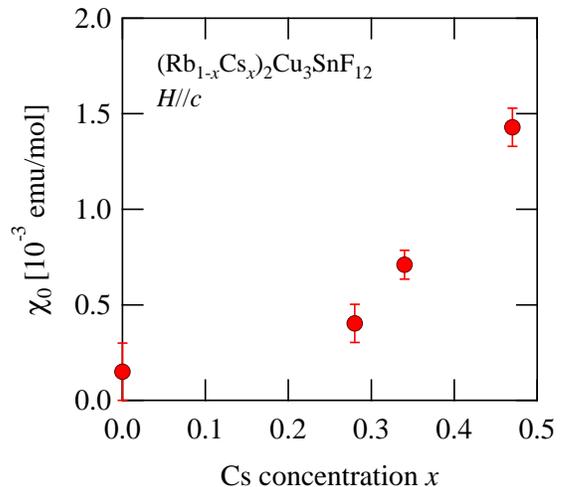}
\end{center}
\caption{
Residual susceptibility ${\chi}_{0}$ for $H\,{\parallel}\,c$ in the disordered ground state for $0\,{\le}\,x\,{\le}\,0.47$. 
} \label{chi_0}
\end{figure}

Here we estimate $x$ dependence of ${\chi}_{0}$ from the magnetization curves of (Rb$_{1-x}$Cs$_{x}$)$_2$Cu$_3$SnF$_{12}$ ($0 \, {\leq} x\, {\leq} \, 0.47$) for $H\,{\parallel}\,c$ at 1.8 K shown in Fig. \ref{field}.
The magnetization data have been corrected for impurity contributions which are assumed to follow the Brillouin function.
The impurity concentration were evaluated to be between 0.4 \% and 0.7 \%.
For $x\, =\, 0$, the highest applied field of 7\,T is smaller than the critical value $H_c=13$ T, where the excitation gap closes.~\cite{Morita_JPSJ2008,Ono_PRB2009}
The magnetization slope below 2\,T is very small but finite, from which ${\chi}_{0}$ is estimated to be ${\chi}_{0}\,{\simeq}\,1\,{\times}\,10^{-4}$ emu/mol. The residual susceptibility ${\chi}_{0}$ for $H\,{\perp}\,c$ is estimated as ${\chi}_{0}\,{\simeq}\,4\,{\times}\,10^{-4}$ emu/mol, which is four times larger than that for $H\,{\parallel}\,c$. The finite ${\chi}_{0}$ is attributed to the small transverse component $D^{\perp}$ of the $\bm D$ vector for the DM interaction.\cite{Fukumoto} 
Figure \ref{chi_0} summarizes the residual susceptibility ${\chi}_{0}$ for $x\,{\leq}\,0.47$ estimated from the magnetization slope. The residual susceptibility ${\chi}_{0}$ is finite for $H\,{\parallel}\,c$ even in the disordered ground state and exhibits a rapid increase with increasing $x$.
For $x\,=\,0.47$, magnetization increases rapidly up to $H_c\,{\sim}\,6.2$ T and increases linearly with increasing magnetic field.
The magnitude of the gap ($\,{\simeq}\,9.8$\,K) for $x\,=\,0.47$ is consistent with that obtained from the low-temperature susceptibility using Eq. (1), as shown below.

Fitting Eq.\,(\ref{Gap}) to the low-temperature susceptibility of Rb$_2$Cu$_3$SnF$_{12}$ for $H\,{\parallel}\,c$ with ${\chi}_{0}\,{\simeq}\,1\,{\times}\,10^{-4}$ emu/mol, we obtain ${\Delta}/k_{\rm B}\,{=}\,28$ K, which is consistent with ${\Delta}/k_{\rm B}\,{=}\,27$ K observed by neutron inelastic scattering.\cite{Matan_Nature_Phys2010,Matan_PRB2014} This guarantees the validity of the present analysis. The $x$ dependence of the excitation gap $\Delta$ obtained by fitting Eq.\,(\ref{Gap}) with ${\chi}_{0}$ shown in Fig. \ref{chi_0} is shown in Fig.\,\ref{phase}. With increasing $x$, $\Delta$ diminishes and vanishes at $x_{\rm c}\,{\simeq}\,0.53$. Because both the excitation gap $\Delta$ and the ordering temperature $T_{\rm N}$ become zero at $x_{\rm c}\,{\simeq}\,0.53$, we can deduce that a quantum phase transition from the disordered state to the ordered state takes place at $x\,{=}\,x_{\rm c}$. Therefore, $x_{\rm c}\,{\simeq}\,0.53$ should be the quantum critical point.




The ground states for the quantum triangular lattice and kagome lattice antiferromagnets with bond randomness and site dilution have been discussed theoretically, and the valence-bond-glass (VBG) phase was argued to be the ground state.\cite{Tarzia_EPL2008,Singh_PRL2010,Watanabe,Kawamura} The VBG phase has finite susceptibility but no long-range ordering. The VBG phase is similar to the Bose glass phase, which emerges in an interacting boson system with random potential\,\cite{Giamarchi,Fisher} and/or in a disordered dimer magnet in a magnetic field.\cite{Oosawa_PRB2002,Shindo,Yamada,Zheludev} The low-temperature magnetic properties characteristic of the VBG have actually been observed in the spatially anisotropic triangular-lattice antiferromagnet Cs$_2$Cu(Br$_{1-x}$Cl$_x$)$_4$.\cite{Ono_JPSJ2005} Because the magnetic susceptibility of (Rb$_{1-x}$Cs$_{x}$)$_2$Cu$_3$SnF$_{12}$ in the ground state is finite for $0\,{<}\,x\,{\le}\,0.47$, the ground state appears to be gapless. Hence the gap $\Delta$ shown in Fig.\,\ref{phase} should be the pseudogap. Because the ground state properties for $0\,{<}\,x\,{\le}\,0.47$ are consistent with those for the VBG, we infer that the ground state can be described as the VBG and that the phase transition at $x_{\rm c}\,{\simeq}\,0.53$ corresponds to the transition from the VBG state to the ordered state with the $q\,{=}\,0$ structure. Microscopic measurements are necessary to clarify the nature of the disordered ground state in (Rb$_{1-x}$Cs$_{x}$)$_2$Cu$_3$SnF$_{12}$.


Note that the temperature dependence of the magnetic susceptibility of (Rb$_{1-x}$Cs$_{x}$)$_2$Cu$_3$SnF$_{12}$ for $0\,{<}\,x\,{\le}\,0.47$  is similar to that of the $S\,{=}\,1/2$ kagome-lattice antiferromagnet herbertsmithite, ZnCu$_3$(OH)$_6$Cl$_2$, extracted from the Knight shift of NMR spectra. \cite{Imai,Olariu} This will give insight into the ground state of herbertsmithite. In an actual sample of herbertsmithite, Cu$^{2+}$ partially substitutes for Zn$^{2+}$.\cite{Bert_PRB2007} Cu$^{2+}$ in an octahedral environment is Jahn-Teller active. Consequently, the substituted Cu$^{2+}$ pushes and pulls the surrounding oxygen ions, which leads to disorder in the oxygen position. Because the oxygen mediates the superexchange interaction in the kagome layer and the superexchange interaction is sensitive to the bond angle of Cu$^{2+}$$-$\,O$^{2-}$$-$\,Cu$^{2+}$, the exchange interaction in the kagome layer is considered to be nonuniform, as in (Rb$_{1-x}$Cs$_{x}$)$_2$Cu$_3$SnF$_{12}$. Therefore, it is plausible that the ground state of the actual sample of herbertsmithite is similar to that in the disordered state of (Rb$_{1-x}$Cs$_{x}$)$_2$Cu$_3$SnF$_{12}$.
 

\section{Conclusion}

We have systematically investigated the variation in the exchange interactions and the magnetic ground states in the spatially anisotropic kagome-lattice antiferromagnet (Rb$_{1-x}$Cs$_{x}$)$_2$Cu$_3$SnF$_{12}$ by magnetic susceptibility measurements. It was found that the  four sorts of nearest-neighbor exchange interactions tend to become more uniform with increasing cesium concentration $x$. We have found that a quantum phase transition from the disordered state to the ordered state occurs at $x_{\rm c}\,{\simeq}\,0.53$. The disordered state for $x\,{<}\,x_{\rm c}$ has finite magnetic susceptibility and a pseudogap $\Delta$, which decreases with increasing $x$ and vanishes at $x_{\rm c}$. The disordered state is concluded to be the valence-bond-glass state.


\section*{ACKNOWLEDGMENTS}
We are indebted to S. Hirata for his technical advice on the exact diagonalization. We express our profound gratitude to Y. Ohtsuka for assistance with ICP-MS analysis. K. K. acknowledges the financial support from the Center of Excellence Program by MEXT, Japan through the ``Nanoscience and Quantum Physics" Project of the Tokyo Institute of Technology. This work was supported by Grant-in-Aid for Scientific Research (A) (Grants No. 23244072 and No. 26247058) and Grant-in-Aid for Young Scientists (B) (Grant No. 26800181) from the Japan Society for the Promotion of Science.


\begin{thebibliography}{25}
\expandafter\ifx\csname natexlab\endcsname\relax\def\natexlab#1{#1}\fi \expandafter\ifx\csname
bibnamefont\endcsname\relax
  \def\bibnamefont#1{#1}\fi
\expandafter\ifx\csname bibfnamefont\endcsname\relax
  \def\bibfnamefont#1{#1}\fi
\expandafter\ifx\csname citenamefont\endcsname\relax
  \def\citenamefont#1{#1}\fi
\expandafter\ifx\csname url\endcsname\relax
  \def\url#1{\texttt{#1}}\fi
\expandafter\ifx\csname urlprefix\endcsname\relax\def\urlprefix{URL }\fi
\providecommand{\bibinfo}[2]{#2} \providecommand{\eprint}[2][]{\url{#2}}

\bibitem{Balents_Nature2010}
L. Balents, 
Nature (London) {\bf 464}, 199 (2010).

\bibitem{Chubukov} 
A. Chubukov, Phys. Rev. Lett. \textbf{69}, 832 (1992).

\bibitem{Sachdev} 
S. Sachdev, Phys. Rev. B \textbf{45}, 12377 (1992).

\bibitem{Lawler_PRL2008}
M. J. Lawler, L. Fritz, Y. B. Kim, and S. Sachdev, 
Phys. Rev. Lett. {\bf 100}, 187201 (2008).

\bibitem{Singh}  
R. R. P. Singh and D. A. Huse, Phys. Rev. B \textbf{77}, 144415 (2008).

\bibitem{Sorensen_PRB2009}
E. S. S\o rensen, M. J. Lawler, and Y. B. Kim,
 Phys. Rev. B {\bf 79}, 174403 (2009).

\bibitem{Evenbly} 
G. Evenbly and G. Vidal, Phys. Rev. Lett. \textbf{104}, 187203 (2010).

\bibitem{Hwang1} 
K. Hwang, Y. B. Kim, J. Yu, and K. Park: Phys. Rev. B \textbf{84}, 205133 (2011).
 
\bibitem{Hastings_PRB2000}
M. B. Hastings,
 Phys. Rev. B {\bf 63}, 014413 (2000).

\bibitem{Wang} 
F. Wang and A. Vishwanath, Phys. Rev. B \textbf{74}, 174423 (2006).%
                
\bibitem{Hermele} 
M. Hermele, Y. Ran, P. A. Lee, and X.-G. Wen, Phys. Rev. B \textbf{77}, 224413 (2008).

 \bibitem{Yan_Science2011}
 S. Yan, D. A. Huse, and S. R. White, 
 Science {\bf 332}, 1173 (2011).
 
\bibitem{Waldmann_EPJB1998}
Ch. Waldtmann, H. U. Everts, B. Bernu, C. Lhuillier, P. Sindzingre, P. Lecheminant, and L. Pierre, 
 Eur. Phys. J. B {\bf 2}, 501 (1998).
 
 \bibitem{Nakano_JPSJ2011}
 H. Nakano and T. Sakai,
 J. Phys. Soc. Jpn. {\bf 80}, 053704 (2011).
 
\bibitem{Depenbrock}  
S. Depenbrock, I. P. McCulloch, and U. Schollw\"ock, Phys. Rev. Lett. {\bf109}, 067201 (2012).
 
 \bibitem{Nishimoto_NatComm2013}
S. Nishimoto, N. Shibata, and C. Hotta,
Nat. Commun. {\bf 4}, 2287 (2013).

 \bibitem{Hiroi_JPSJ2001}
Z. Hiroi, M. Hanawa, N. Kobayashi, M. Nohara, H. Takagi, Y. Kato, and M. Takigawa,
J. Phys. Soc. Jpn. {\bf 70}, 3377 (2001).

 \bibitem{Yoshida_NatComm2012}
H. Yoshida, J. Yamaura, M. Isobe, Y. Okamoto, G. J. Nilsen and Z. Hiroi,
Nat. Commun. {\bf 3}, 860 (2012).

 \bibitem{Yoshida_JPSJ2013}
M. Yoshida, Y. Okamoto, M. Takigawa, and Z. Hiroi,
J. Phys. Soc. Jpn. {\bf 82}, 013702 (2013).

 \bibitem{Okamoto_JPSJ2009}
Y. Okamoto, H. Yoshida, and Z. Hiroi,
J. Phys. Soc. Jpn. {\bf 78}, 033701 (2009).

\bibitem{Shores_JACS2005}
M. P. Shores, E. A. Nytko, B. M. Bartlett, and D. G. Nocera, 
J. Am. Chem. Soc. {\bf 127}, 13462 (2005).

\bibitem{Mendels_PRL2007}
P. Mendels, F. Bert, M. A. de Vries, A. Olariu, A. Harrison, F. Duc, J. C. Trombe, J. S. Lord, A. Amato, and C. Baines,
Phys. Rev. Lett. {\bf 98}, 077204 (2007).

\bibitem{Helton}
J. S. Helton, K. Matan, M. P. Shores, E. A. Nytko, B. M. Bartlett, Y. Yoshida, Y. Takano, A. Suslov, Y. Qiu, J. -H. Chung, D. G. Nocera, and Y. S. Lee, Phys. Rev. Lett. \textbf{98}, 107204 (2007).

\bibitem{Mueller} 
M. M\"{u}ller and B. G. M\"{u}ller, Z. Anorg. Allg. Chem. \textbf{621}, 993 (1995).

\bibitem{Janson}
O. Janson, J. Richter, P. Sindzingre, and H. Rosner, Phys. Rev. B \textbf{82}, 104434 (2010).


\bibitem{Bert_PRB2007}
F. Bert, S. Nakamae, F. Ladieu, D. L'H\^{o}te, P. Bonville, F. Duc, J.-C. Trombe, and P. Mendels,
Phys. Rev. B {\bf 76}, 132411 (2007).

\bibitem{Reisinger}
S. A. Reisinger, C. C. Tang, S. P. Thompson, F. D. Morrison, and P. Lightfoot, Chem. Mater. \textbf{23}, 4234 (2011).

\bibitem{Morita_JPSJ2008}
K. Morita, M. Yano, T. Ono, H. Tanaka, K. Fujii, H. Uekusa, Y. Narumi, and K. Kindo,
J. Phys. Soc. Jpn. {\bf 77}, 043707 (2008).

\bibitem{Ono_PRB2009}
T. Ono, K. Morita, M. Yano, H. Tanaka, K. Fujii, H. Uekusa, Y. Narumi, and K. Kindo,
Phys. Rev. B {\bf 79}, 174407 (2009).

\bibitem{Downie} L. J. Downie, C. Black, E. I. Ardashnikova, C. C. Tang, A. N. Vasiliev,
A. N. Golovanov, P. S. Berdonosov, V. A. Dolgikh, and P. Lightfoot, CrystEngComm. \textbf{16}, 7419 (2014).

\bibitem{Ono_JPSJ2014}
 T. Ono, K. Matan, Y. Nambu, T. J. Sato, K. Katayama, S. Hirata, and H. Tanaka,
J. Phys. Soc. Jpn. {\bf 83}, 043701 (2014).

\bibitem{Matan_Nature_Phys2010}
K. Matan, T. Ono, Y. Fukumoto, T. J. Sato, J. Yamaura, M. Yano, K. Morita, and H. Tanaka, 
Nat. Phys. {\bf 6}, 865 (2010).

\bibitem{Matan_PRB2014}
K. Matan, Y. Nambu, Y. Zhao, T. J. Sato, Y. Fukumoto, T. Ono, H. Tanaka, C. Broholm, A. Podlesnyak, and G. Ehlers,
Phys. Rev. B {\bf 89}, 024414 (2014).

\bibitem{Grbic_PRL2013}
M. S. Grbi\'c, S. Kr\"amer, C. Berthier, F. Trousselet, O. C\'epas, H. Tanaka, and M. Horvati\'c,
Phys. Rev. Lett. {\bf 110}, 247203 (2013).

\bibitem{Yang_PRB2009} B. J. Yang and Y. B. Kim, Phys. Rev. B {\bf79}, 224417 (2009).

\bibitem{Khatami_PRB2011} E. Khatami, R. R. P. Singh, and M. Rigol, Phys. Rev. B {\bf84}, 224411 (2011).

\bibitem{Cepas} 
O. C\'{e}pas, C. M. Fong, P. W. Leung, and C. Lhuillier, Phys. Rev. B \textbf{78}, 140405(R) (2008).

\bibitem{dCP}
J. des Cloizeaux and J. J. Pearson, Phys. Rev. \textbf{128}, 2131 (1962).

\bibitem{Endoh}
Y. Endoh, G. Shirane, R. Birgeneau, P. Richards, and S. Holt, Phys. Rev. Lett. \textbf{32}, 170 (1974).

\bibitem{Igarashi} 
J. Igarashi, Phys. Rev. B \textbf{46}, 10763 (1992).

\bibitem{Singh2} 
R. R. P. Singh, and M. P. Gelfand, Phys. Rev. B \textbf{52}, R15695 (1995).

\bibitem{Ronnow} H. M. R{\o}nnow, D. F. McMorrow, R. Coldea, A. Harrison, I. D. Youngson, T. G. Perring, G. Aeppli, O. Sylju{\aa}sen, K. Lefmann, and C. Rischel, Phys. Rev. Lett. \textbf{87}, 037202 (2001).

\bibitem{Hwang2}
K. Hwang, K. Park, and Y. B. Kim, Phys. Rev. B \textbf{86}, 214407 (2012).

\bibitem{Selwood_Wiley1956}
 P. W. Selwood, Magnetochemistry (Wiley-Interscience, New York,
1956) 2nd ed., Chap. 2, p. 78.

\bibitem{Sasaki_JPSJ1995}
S. Sasaki, N. Narita, and I. Yamada,
J. Phys. Soc. Jpn. {\bf 64}, 2701 (1995).

\bibitem{Troyer_PRB1994}
M. Troyer, H. Tsunetsugu, and D. W\"{u}rtz, 
Phys. Rev. B {\bf 50}, 13515 (1994).

\bibitem{Stone_PRB2001}
M. B. Stone, I. Zaliznyak, D. H. Reich, and C. Broholm, 
Phys. Rev. B {\bf 64}, 144405 (2001).

\bibitem{Fukumoto} K. Sasao, N. Kunisada, and Y. Fukumoto, unpublished result.
K. Sasao, N. Kunisada, and Y. Fukumoto, Meeting Abst. Phys. Soc. Jpn. {\bf 65(2-3)}, 442 (2010). (in Japanese) 


\bibitem{Tarzia_EPL2008}
M. Tarzia and G. Biroli,
Europhys. Lett. {\bf 82}, 67008 (2008).

\bibitem{Singh_PRL2010}
R. R. P. Singh,
Phys. Rev. Lett. {\bf 104}, 177203 (2010).

\bibitem{Watanabe}
K. Watanabe, H. Kawamura, H. Nakano, and T. Sakai, J. Phys. Soc. Jpn. \textbf{83}, 034714 (2014).

\bibitem{Kawamura} H. Kawamura, K. Watanabe, and T. Shimokawa, J. Phys. Soc. Jpn. \textbf{83}, 103704 (2014).

\bibitem{Giamarchi} T. Giamarchi and H. J. Schulz, Phys. Rev. B \textbf{37}, 325 (1988).

\bibitem{Fisher} M. P. A. Fisher, P. B. Weichman, G. Grinstein, and D. S. Fisher, Phys. Rev. B \textbf{40}, 546 (1989).

\bibitem{Oosawa_PRB2002}
A. Oosawa and H. Tanaka, Phys. Rev. B {\bf 65}, 184437 (2002).

\bibitem{Shindo} Y. Shindo and H. Tanaka, J. Phys. Soc. Jpn. \textbf{73}, 2642 (2004).

\bibitem{Yamada} F. Yamada, H. Tanaka, T. Ono, and H. Nojiri, Phys. Rev. B \textbf{83}, 020409(R) (2011).

\bibitem{Zheludev} A. Zheludev and T. Roscilde, C. R. Phys. \textbf{14}, 740 (2013).


\bibitem{Ono_JPSJ2005} 
T. Ono, H. Tanaka, T. Nakagomi, O. Kolomiyets, H. Mitamura, F. Ishikawa, T. Goto, K. Nakajima, A. Oosawa, Y. Koike, K. Kakurai, J. Klenke, P. Smeibidle, M. Meisner, and H. A. Katori, J. Phys. Soc. Jpn. \textbf{74} (Suppl.), 135 (2005).

\bibitem{Imai} T. Imai, E. A. Nytko, B. M. Bartlett, M. P. Shores, and D. G. Nocera, Phys. Rev. Lett. \textbf{100}, 077203 (2008).

\bibitem{Olariu} 
A. Olariu, P. Mendels, F. Bert, F. Duc, J. C. Trombe, M. A. de Vries, and A. Harrison, Phys. Rev. Lett. \textbf{100}, 087202 (2008).



\end{thebibliography}
\end{document}